# Radiobiological Characterization of Clinical Proton and Carbon-Ion Beams

*P. Scalliet and J. Gueulette*
Université Catholique de Louvain, Cliniques Universitaires Saint Luc, 10, avenue Hippocrate, 1200 Brussels, Belgium

**Abstract**
Electromagnetic radiation (photons) or particle beam (protons or heavy ions) have similar biological effects, i.e. damage to human cell DNA that eventually leads to cell death if not correctly repaired. The biological effects at the level of organs or organisms are explained by a progressive depletion of constitutive cells; below a given threshold, cell division is no longer sufficient to compensate for cell loss, up to a point where the entire organism (or organ) breaks down. The quantitative aspects of the biological effects are modulated by the microscopic distribution of energy deposits along the beam or particle tracks. In particular, the ionization density, i.e. the amount of energy deposited by unit path length (measured in keV/μm), has an influence on the biological effectiveness, i.e. the amount of damage per energy unit deposited (measured in gray or Gy, equivalent to 1 joule/kg). The ionization density is usually represented by the Linear Energy Transfer or LET, also expressed in keV/μm. Photon beams (X-rays, g-rays) are low-LET radiation, with a sparsely ionising characteristic. Particle beams have a higher LET, with a more dense distribution of energy deposits along the particle tracks. Protons are intermediary, with a LET larger than the photon one, but still belong to the 'radiobiological' group of low LET. The higher the ionization density, the higher the biological effectiveness per unit of dose. When comparing various radiation qualities, it appears that the ionization density is relatively homogeneous along photon tracks, whereas it strongly varies along particular tracks (protons, heavy ions). In the first instance, the biological effectiveness is proportional to the TEL, itself dependant on the particle beam energy. So, when the LET of a particle beam is increased, its biological effectiveness increases in proportion. Secondly, a low-energy beam (f.i. 4 MeV a rays) has a higher LET than a high-energy beam (f.i. 200 MeV a rays). As particle beams continuously loose their energy through their successive interactions with the irradiated medium, it ensues that the LET slowly increases along the beam path, down to a point where all energy has been imparted and the beam stops. Therefore, the biological effectiveness is not homogeneous along the beam path (like with low-LET radiation), with a strong reinforcement at the end of the particle tracks (in the Bragg peak). The modelization of the clinical effects of particle beams is therefore very challenging, as a variable biological weighting function needs to be incorporated in the planning process to account for the increase in biological effectiveness with the progressive loss of beam energy.

# 1 Introduction

The performance of radiotherapy can be improved in two separate ways: (a) improvement of the ballistic selectivity (increasing the dose to the tumour while reducing the exposure of normal tissue) and (b) improvement of the biological effectiveness of the radiation (using radiation with a higher relative biological effectiveness, RBE). Clinical proton beams are ballistically superior but biologically equivalent to X-rays (and gamma rays), while carbon-ion beams are both ballistically and biologically more efficient than X-rays.

From the biological point of view, the effect of radiation on living material is mainly due to DNA damage and its consequences: DNA disruption, loss of genetic information, incapacitation of vital genes, and, eventually, cell death. Indeed, severe DNA damage rapidly induces cell apoptosis (a sophisticated mechanism of auto-destruction) if the DNA is not correctly repaired in time (a few hours). It may, however, happen that some cells manage to survive despite severe DNA damage, but they often do so with some amount of 'misrepair', i.e., incorrectly repaired DNA damage with a change in the information sequence (gene inactivation, gene promotion, etc.).

Cells that survive severe DNA damage are rare, but can be dangerous if they harbour gene alterations that may lead to cancer (loss of proliferation regulation, loss of apoptosis, and tissue invasion and colonization).

The sequence of events leading to cell death can be summarized in the following way:

– Energy is deposited in DNA (by primary and secondary electrons) along radiation tracks, in consecutive ionization events. In fact, what happens is an 'exchange' of energy between the radiation and peripheral electrons of the atoms constituting the DNA, causing electrons to break loose from atoms and collide further with other, neighbouring atoms. Often, a cluster of ionization arises from the accumulation of the primary energy exchange event and the interactions of secondary electrons.

– The appearance of positive charges in the DNA molecule causes a rearrangement or, worse, a complete disruption of the molecular structure if the energy imparted exceeds the binding energy of the atoms. The sequence of information coded in the DNA strands is therefore interrupted by a break.

– DNA rearrangement follows detection of damage. This process is amazingly fast: induction of repair enzymes starts within minutes after DNA damage.

– Repair induction (enzyme synthesis) occurs in proportion to the amount of damage, though only up to a certain point, as massive damage tends to saturate the repair mechanisms. Repair usually takes 4 to 6 hours. Slight damage can be repaired faster; severe damage takes longer. Owing to the dual structure of DNA, i.e., a structure of two long strands that mirror each other, a single-strand break is easily repaired, but a double-strand break is not. The repair mechanisms excise damaged DNA sequences and rebuild intact DNA by reading the 'mirror' strand. If both strands are damaged, the danger of a faulty repair (misrepair), i.e., a repair resulting in DNA with modified information, is larger.

– The cell cycle is arrested at the same time as repair induction, in order to 'lend' sufficient time for repair before the next cell division is triggered.

– Along with DNA repair, apoptosis is also triggered. This might look illogical but, in fact, it is a strong protective mechanism against misrepair. Apoptosis is a stepwise process, each step being reversible up to a certain 'no-return' point at which the process becomes unstoppable. This point is reached after a definite time period. If at that point the repair is not finished (because the damage is too dense), then apoptosis proceeds until cell death. Conversely, if the repair is finished before that point (because the damage is limited), then apoptosis is stopped and the cell survives. How does this mechanism protect cells? In fact, it does not protect individual cells, but it protects the information conveyed by a cell population, by eliminating severely damaged cells that are at risk of misrepair and corruption of the information in the DNA.[1]

- The cell dies (as a result of apoptosis or misrepair) or survives (with either adequate or inadequate DNA repair).
- Tissue failure (in the case of normal tissue) occurs if enough cells have been destroyed, or cancer cure is achieved (in the case of cancer) if all cancer cells have been destroyed.

## 2    Density of ionization and microdosimetry

Whether the damage to DNA is light or severe depends in the first instance on the amount of energy dissipated in the molecule (the 'dose'), but it also depends on the density of ionizing tracks crossing the molecule. A few dense tracks are biologically more effective than several sparsely ionizing tracks. Therefore, smaller doses with 'dense' tracks are as effective in killing cells as larger doses with less dense tracks.

This density is related to the amount of energy per unit track length and to the distance between consecutive energy deposition events along the track of the particle (a photon in the case of X-rays, or a proton or carbon ion in hadron therapy); more specifically, it is related to the number of energy deposition or ionization events that occur in the short diameter of the DNA (a few nanometres). Densely ionizing radiation usually deposits enough energy to inactivate a cell in one single track, whereas sparsely ionizing radiation requires the cooperation of several tracks, each depositing a small amount of energy insufficient to kill a cell, to achieve the same result (Fig. 1).

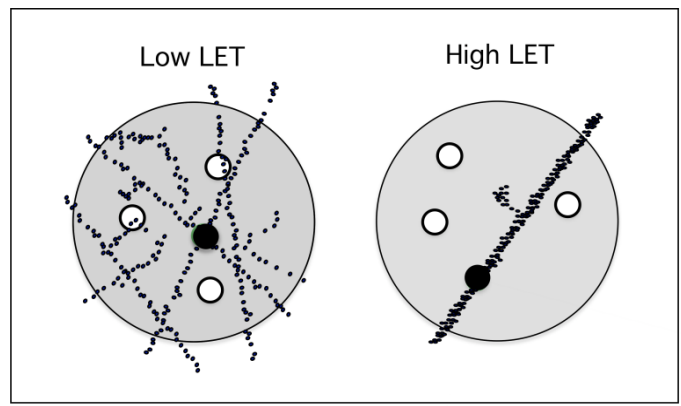

**Fig. 1:** Schematic representation of particle tracks for low-LET (left) and high-LET (right) radiation [1, 2]. For low-LET radiation, the inactivation of a radiosensitive target requires the conjunction of several tracks, whereas for high-LET radiation the impact of a single track is always fatal (closed circles).

The linear energy transfer (LET), a quantity expressed in keV/µm of particle track, measures the density of ionization per unit length along radiation tracks. Types of radiation can be sorted by their LET, with a customary distinction between low-LET (<10–20 keV/µm) and high-LET (>20 keV/µm) radiation.

Energy deposition in DNA is a quantized or random event, sometimes important, sometimes not. But the maximum energy that can be imparted in a single interaction depends directly on the LET. Low-LET radiation therefore very seldom kills cells with a 'single hit', whereas this is very common with high-LET radiation.

---

[1] This explains why embryos exposed to radiation most often do not survive. Indeed, inheritable DNA mutations have not been observed in the Hiroshima and Nagasaki survivors.

One particular dosimetric method allows individual energy deposition events to be measured in the eV range. This method is called 'microdosimetry', as its purpose is to describe energy exchange at the molecular level. In short, the overall concept is to miniaturize a dosimeter and to expose it to an extremely low particle fluence in order to register interactions separately. These interactions are measured by collecting electric charges created in a counter whose volume is artificially reduced by lowering the gas pressure in the measurement chamber. A very low pressure of a tissue-equivalent gas mimics a very small volume at normal atmospheric pressure, in the range of cubic micrometres. Nowadays, most microdosimetry is no longer done physically but done 'in silico' using Monte Carlo simulation methods.

Irradiating at low fluence and integrating all energy deposition events in a single graph yields a 'microdosimetric spectrum' specific to the radiation or particle type tested (Fig. 2). Small variations in the particle energy are reflected in small variations in the microdosimetric spectrum. In turn, variations in the microdosimetric spectrum illustrate differences in biological effectiveness, i.e., in the proportion of cells irreversibly damaged when it comes to cell kill.

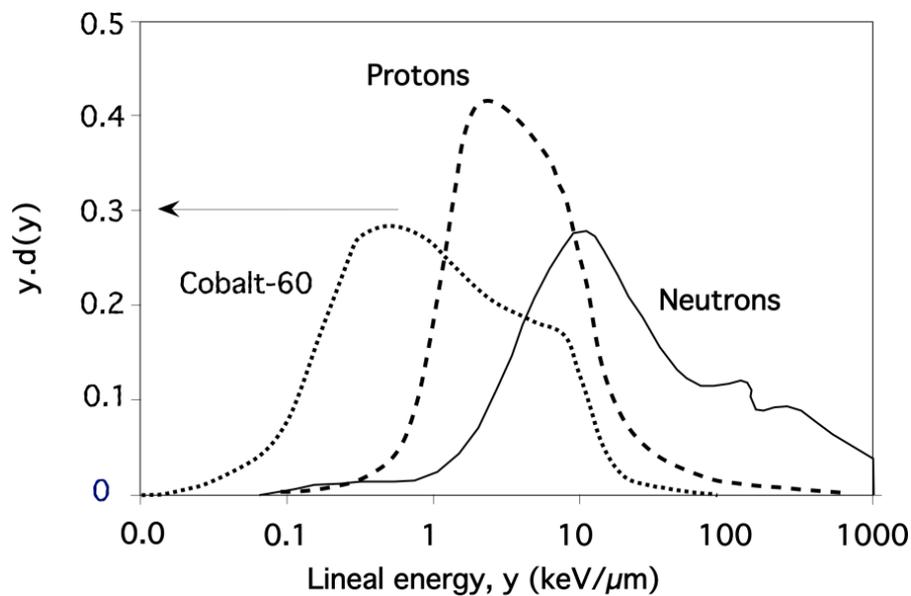

**Fig. 2:** Comparison of microdosimetric spectra of $y.d(y)$ vs. $y$ obtained for cobalt-60 gamma rays, 65 MeV protons, and p(65) + Be neutrons [3]; $y$ is the lineal energy and $d(y)$ is the probability density of the absorbed dose with respect to $y$. For cobalt-60 gamma rays, the maximum $y.d(y)$ values occur at about 0.3 keV/μm. For protons and neutrons, the maxima are observed at about 3 keV/μm and 10 keV/μm, respectively.

## 3    Relative biological effectiveness

Again, minute energy deposition events are unable to damage DNA in a significant manner, whereas massive energy deposition events invariably kill the cell. When different types of radiation are compared (e.g., X-rays, neutrons, and alpha particles), the relationship between dose and cell survival shows that for the same amount of energy dissipated (i.e., the same radiation dose), the number of cells killed increases with the LET (Fig. 3).

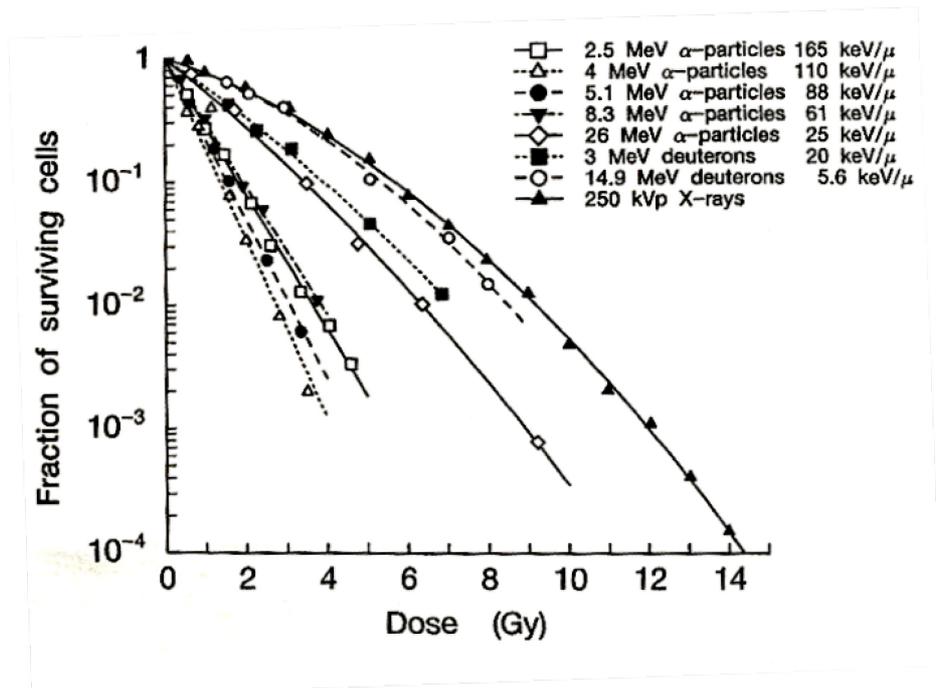

**Fig. 3:** Survival curves for cells exposed to radiation of different LET. The slopes of the curves become steeper as the LET of the radiation increases. The bending of the curves (i.e., the initial shoulder) is also progressively reduced. (Redrawn from Ref. [4].)

Conversely, for an identical cell kill level (known as an isoeffect), the dose required decreases as the LET increases. The ratio of the low-LET dose to the high-LET dose required for an isoeffect is called the RBE. It tends to increase with LET up to a maximum value, depending on the isoeffect considered (Fig. 5). Above this maximum value, the RBE decreases as the ionization density becomes very large, and most of the energy is wasted (an overkill phenomenon).

The RBE is a dimensionless quantity (as it is a ratio of doses) that compares the biological effectiveness of a given type of radiation with another type taken as a reference, usually cobalt-60 gamma rays. Thus, when cobalt-60 radiation is compared with itself, the RBE is 1. Up to an LET of around 20 keV/μm, the RBE remains stable at 1. Above this LET value, the RBE increases rapidly to a maximum value at around 100 keV/μm.

RBE values depend on the isoeffect level chosen for the comparison of radiation beams. Small radiation doses tend to increase the RBE, since at low doses, low-LET radiation is very ineffective in killing cells, whereas high-LET radiation is quite effective in doing so. At higher doses, low-LET radiation becomes more lethal and the difference in effect between low- and high-LET radiation becomes smaller. At very high doses, the RBE reaches a stable value that no longer depends on the dose (Fig. 4). The RBE also depends on the biological system under consideration.

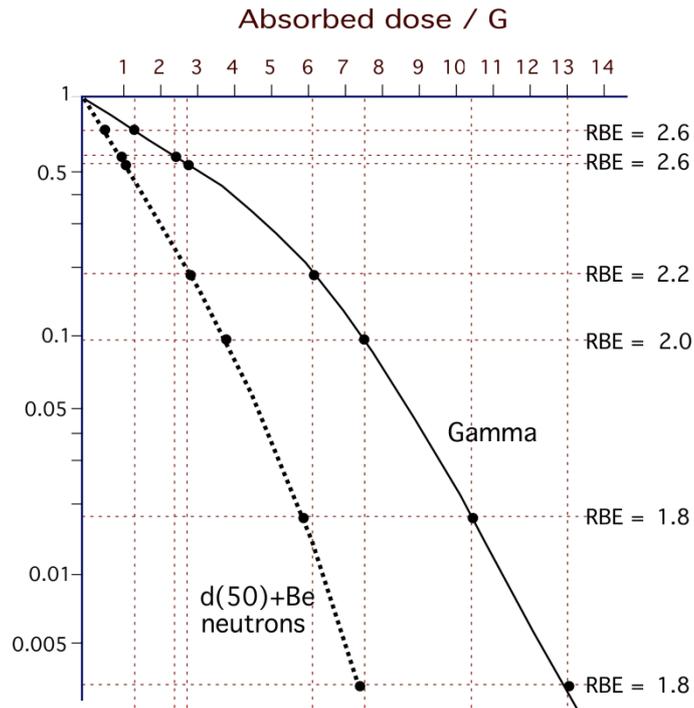

**Fig. 4:** Survival curves for intestinal crypt cells irradiated with neutrons or cobalt-60 gamma rays. The RBE is particularly variable in the initial part of the curves (i.e., for small doses), where it reaches its highest value. This variation is mainly due to the bending (i.e., the shoulder) of the gamma-ray curve. As the dose increases, the RBE stabilizes progressively, tending towards a minimum.

## 4     Time factor and repair

So, it is the spatial structure of the ionization events that characterizes the various types of radiation beams. If dense ionization crosses DNA, it will invariably destroy it beyond any possibility of repair. Too much information is lost in the event. Carbon-ion beams belong to the class of radiation that causes such dense ionization, i.e., high-LET radiation.

Low-LET radiation, conversely, only kills cells by the cooperation of several tracks that occur together spatially and in time: spatially to 'build up' damage at a particular molecular site, and in time because consecutive 'hits' must occur before the previous 'hit' has been repaired. For this reason, lowering the dose rate of the irradiation sharply decreases the biological effectiveness, as more time is available for the repair of damage before the next ionization takes place. Conversely, lowering the dose rate with high-LET radiation does not alter the biological effectiveness much, since a single hit is sufficient to inactivate a cell.

Another way to decrease the biological effectiveness of low-LET radiation is to deliver the dose in several small fractions, separated by sufficient time for DNA repair. In this case, spatial cooperation does not work, since small amounts of DNA damage are repaired between consecutive fractions. The longer the time between fractions, the more thorough the repair of the DNA. Again, this does not influence the biological effectiveness of high-LET radiation, as one single hit in the DNA is usually sufficient to kill the cell, without the need for spatial cooperation.

## 5     The effect of oxygen

It was observed early in the history of radiobiology that the radiosensitivity of cells depends on the partial pressure of oxygen in the immediate environment. When oxygen is present at normal atmospheric concentration, the radiosensitivity is at its highest. Lowering the partial pressure of oxygen progressively decreases the radiosensitivity, by a factor of that reaches 3 when oxygen is absent (or nearly absent) (Fig. 5).

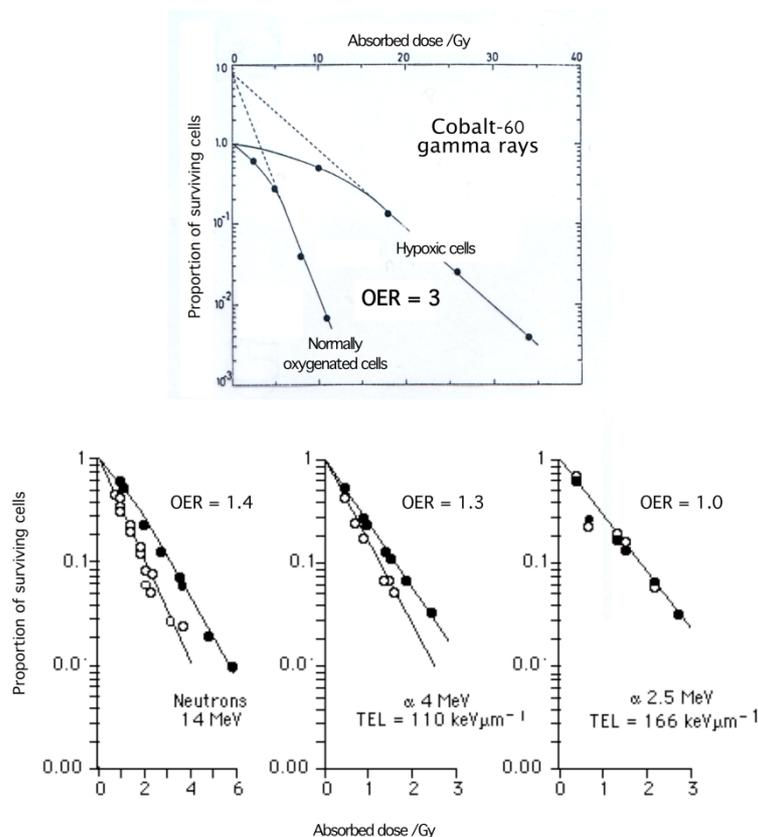

**Fig. 5:** Comparison of the influence of partial pressure of oxygen for radiation with different LETs. OER (Oxygen Enhancement Ratio) equals 3 for low-LET radiation; this value seems to be independent of the dose level and, to a certain extent, of the biological system. The OER value is thus commonly interpreted as a 'scale factor'. The OER value decreases as the LET increases, down to 1 for very high-LET particles. Redrawn from Refs. [4, 5].

Obviously, no aerobic cell (as are human cells f.i.) is viable for more than a few minutes without any oxygen supply. But at a low partial pressure, some survival, often in a quiescent state, remains possible for hypoxic cells, which can then escape the effect of radiation doses that would otherwise be lethal.

In oncology, the blood supply of cancerous masses is commonly deficient, and large regions of hypoxia exist in virtually all tumours. This has proven to be a problem of importance in radiotherapy, and ways to overcome hypoxic radioresistance have been developed (see Section 7).

Oxygen can thus be considered as a 'natural' radiosensitizer. Without entering too much into details, we can say that the presence of oxygen 'fixes' damage in the DNA. What actually happens is that with low-LET radiation, most of the damage is absorbed by water, in the close vicinity of the DNA molecule. In physics terms, the DNA itself has a very small 'cross-section' for X-rays, and most of the damage to it is created by radiolysis of water, which creates free radicals, which in turn interact secondarily with the DNA molecule. The life-span of these free radicals and their ability to migrate some distance is influenced by the presence or absence of oxygen. It is said that oxygen is needed to 'fix' the damage ('fix' in the sense of fixation, not of repair). In its absence, the free radicals are less toxic to the DNA.

This influence of the partial pressure of oxygen is at its highest with low-LET radiation. As the LET of the radiation is progressively increased above 20 keV/μm, the sensitizing effect of oxygen progressively disappears because DNA damage is now usually the consequence of a direct, very dense hit on the molecule, whose fixation no longer requires the presence of oxygen. This is what makes carbon-ion beams clinically so attractive, since their LET is in the range where no oxygen is required to 'fix' the lethal DNA damage. Indeed, carbon ion therapy is advocated for the treatment of cancer types in which a large hypoxic component is suspected.

## 6    Cell cycle and cell division

Cell division is a lengthy and subtle process in which the cell duplicates its DNA before starting to physically divide. The structure of DNA allows this process to be precise and fail-safe (indeed, this is a condition for life). By duplicating each of the two strands constituting the DNA molecule, the cell doubles its set of chromosomes. The two sets then migrate in opposite directions, and the cell is cleaved between them.

DNA synthesis relies on a set of specific enzymes that 'gently' separate the two DNA strands and synthesize a new copy on each of them. The end result is two identical DNA molecules. The same enzymes are mobilized in the case of accidental DNA damage during the lifetime of a cell; they excise the damaged DNA section and then resynthesize the missing part of the DNA, using the other strand as a template for the exact restitution of the coded information.

The synthesis and repair enzymes normally have a very low concentration in the nucleus, at times not close to cell division. But any damage will trigger enzyme synthesis (in a matter of minutes) in order for repair to proceed efficiently.

Conversely, during cell division, the DNA synthesis enzymes are at their maximum concentration in the nucleus. If radiation damage is inflicted during cell division, the cell tends to be more resistant as the nucleus is already saturated with all of the enzymes needed for repair, especially at the end of DNA synthesis, when the enzymes are no longer required for duplication and therefore are free for binding to any new substrate (damaged DNA, for instance). When DNA synthesis is finished, the cell remains quiescent for some time (called the G2 phase), just before physically dividing. At that point the cell is at its most vulnerable to radiation damage, since repair is less effective in the presence of all the rearrangement needed for cell division.

The variation in the sensitivity of cells with the division cycle is more pronounced with low-LET radiation, quite logically, since repair plays an important role in the end result of radiation exposure. In contrast, the sensitivity to high-LET radiation damage is independent of the cell cycle, since repair plays no or only a very minor role in the end result (Fig. 6).

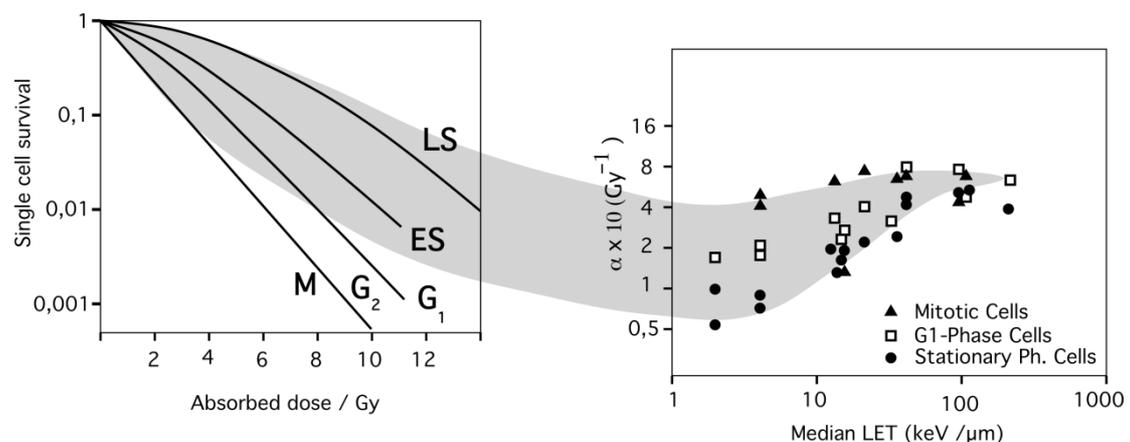

**Fig. 6:** Left panel: survival curves for cells irradiated in different phases of their mitotic cycle. Right panel: variation of the slope of the curves (the parameter $\alpha$) as a function of LET. The gap between the slopes of the various curves (i.e., the difference in radiosensitivity) gradually becomes smaller as the LET increases.

# 7 Biological weighting function

The preceding considerations about the variations of radiosensitivity and the biological effectiveness of low- and high-LET radiation find quantitative expression in the so-called Biological Weighting Function (BWF), which is obtained by plotting the RBE against the LET (Fig. 7) [6, 7]. BWFs are specific to a given biological effect and given irradiation conditions, so that the peak value (the maximum of the RBE) and its place in the LET range may vary substantially. The main sources of variation are the dose and the oxygenation status of the biological material: the RBE of high-LET radiation with respect to cobalt-60 gamma rays increases as the dose decreases and when the partial pressure of oxygen decreases. Recall here that the higher RBE of carbon ions for hypoxic cells with respect to normally oxygenated cells is one of the main justifications for using these particles in radiotherapy.

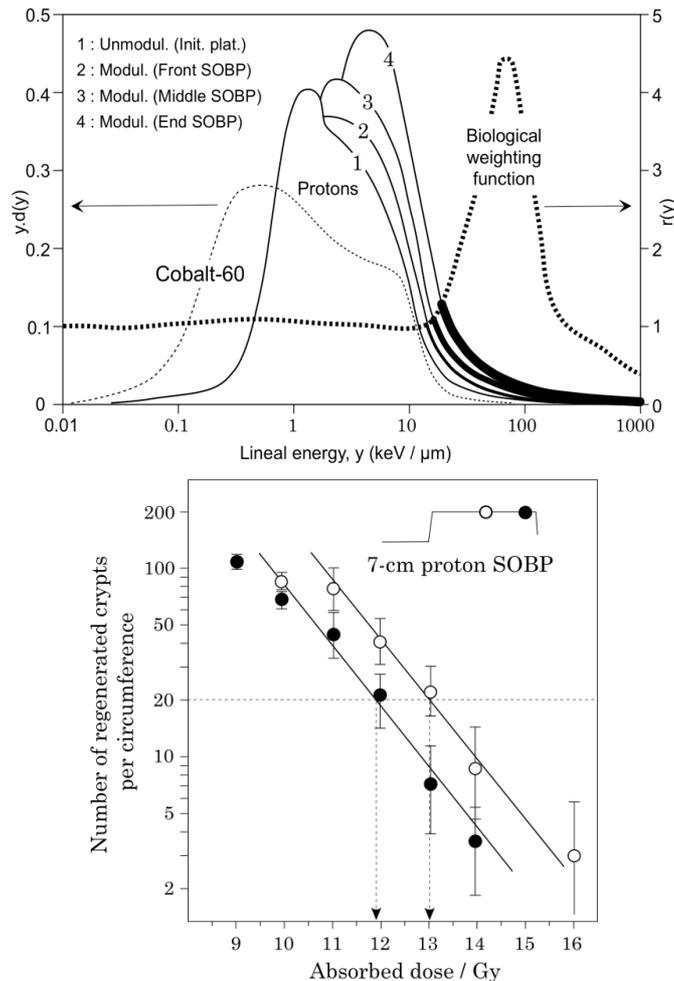

**Fig. 7:** Top: microdosimetric spectra for a 90 MeV energy-modulated proton beam. Measurements at four positions are shown (solid lines). These are compared with cobalt-60 gamma rays (dotted line). The BWF for different LET values has been superimposed (bold dotted line). The proton microdosimetric spectra are shifted towards higher LET values when measured more distally in the SOBP (Spread Out Bragg Peak - bold solid lines). This suggests that the proton RBE would also increase with depth. Bottom: dose–effect relationships for crypt regeneration in mice after irradiation in a single fraction with a 200 MeV energy-modulated proton beam at iThemba Labs, South Africa. The beam was modulated to produce a 7 cm SOBP. The open and closed circles correspond to irradiation in the middle and at the end of the SOBP, respectively (see the sketch of the dose profile at the top of the panel). Each point is the average of the readings for four mice. Parallel exponential regression curves were fitted through the points by a weighted least squares method. The error bars correspond to the 95% confidence intervals. In the case shown here, the RBE increases by 9% on moving from the middle to the end of the SOBP.

# 8 Proton beams

The RBE of protons is just above unity, usually around 1.1–1.15, indicating that the fractionation effect still matters with this type of radiation. Proton beams are a characteristic low-LET radiation.

A more precise examination of their microdosimetric spectrum, however, shows that small variations in LET can be observed along the particle paths. At the entrance to the irradiated medium, high-energy protons are sparsely ionizing, and thus typically low-LET. When they are close to the end of their path, at the place where all the remaining kinetic energy will be released (the Bragg peak), the LET increases sharply, though by only a modest amount. But this small change is sufficient to modify the RBE. Therefore, the RBE is not constant over the entire proton path.

The explanation is quite simple: as the protons enter the medium and penetrate deeper and deeper, they progressively release some of their kinetic energy and slow down, until they reach the end of their path. As the speed decreases, the distance between consecutive energy deposition events also decreases; hence, the LET increases, and the RBE increases in turn. Proton beams thus do not have a constant biological effectiveness along their path. But, again, the variation in RBE remains modest.

Precise radiobiological experiments with mice, using a model of intestinal toxicity, have been able to measure these RBE variations at the end of the proton path, by repeating measurements across the small distance covering the end of an extended Bragg peak. Radiobiological data demonstrate that the RBE indeed varies, and in the proportions predicted by the microdosimetric shift in LET.

# 9 Carbon-ion beams

The physics of carbon-ion beams is similar to that of proton beams (a plateau at the entrance point of the beam, followed by a sharp rise in dose at the end of the path, at the Bragg peak), but the mass of carbon ions is much greater. Therefore, carbon-ion beams are high-LET along their entire path, though with a similar pattern to that of protons: the LET is lower near the entrance point of the beam and at its highest at the Bragg peak.

The range of the RBE is close to 2–3 at the Bragg peak, and closer to 1.5–2 at the entrance point. The variation of the RBE along the path of a carbon ion is therefore much larger than for protons, which calls for some adjustment when irradiating under clinical conditions.

Usually, when a tumour is irradiated with X-rays or gamma rays (low-LET radiation), a homogeneous level of dose is delivered to the entire volume, with the intention of delivering an 'isoeffect' to the tumour. No specific attention to the size of the tumour is needed, as the biological effectiveness of low-LET radiation remains the same across the entire irradiated volume (RBE = 1).

This is not the case with carbon-ion beams, since the biological effectiveness increases significantly within the irradiated volume. If a homogeneous 'biological' dose is planned throughout the tumour volume, then some alteration to the 'physical' dose needs to be made, to compensate for the variation in RBE [8]. Precise measurements that illustrate this variation in RBE have been done with the clinical carbon-ion beam at HIMAC (Fig. 8).

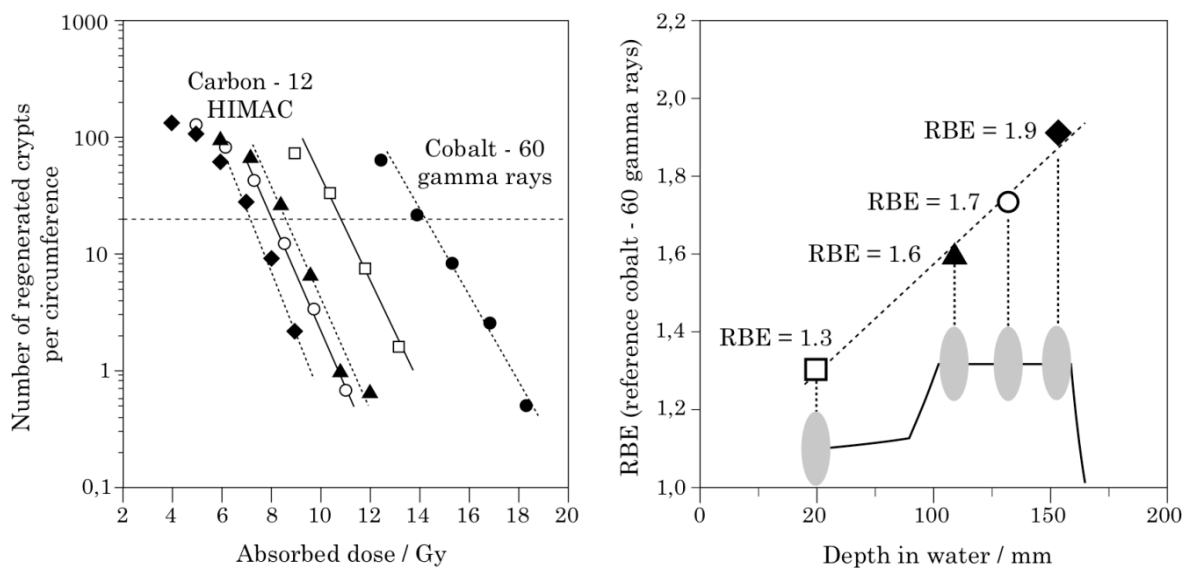

**Fig. 8:** Left panel: dose–effect relationships for intestinal crypt regeneration in mice after irradiation with cobalt-60 gamma rays or carbon-12 ions at the entrance plateau and at different positions in a 6 cm SOBP (the positions are shown in the sketch in the right panel). Right panel: the corresponding RBEs (reference cobalt-60 gamma rays), plotted against the depth, indicate a substantial increase in the RBE. As the irradiations were performed with single high doses, these RBEs are much lower than those for fractionated irradiations, which reach a value of approximately 3 at the end of the SOBP.